\documentclass[showpacs,twocolumn,preprintnumbers,amsmath,amssymb]{revtex4}
\usepackage{amsmath,amsfonts,latexsym,amssymb,graphicx,graphics,epsfig,subfigure,color,makeidx}
\usepackage{xcolor,diagbox}
\usepackage{multirow}
\usepackage[colorlinks,linkcolor=blue,anchorcolor=blue,citecolor=green,urlcolor=blue]{hyperref}
\usepackage{mathrsfs}
\usepackage{amsmath}

\begin{document}
\newcommand \fc{\frac}
\newcommand \lt{\left}
\newcommand \rt{\right}
\newcommand \pd{\partial}
\newcommand \hmn{h_{\mu\nu}}
\newcommand \mn{{\mu\nu}}
\newcommand \dd{\textrm{d}}
\newcommand \e{\textrm{e}}
\newcommand \ii{\textrm{i}}
\newcommand \tcr{\textcolor{red}}
\newcommand{\PR}[1]{\ensuremath{\left[#1\right]}} 
\newcommand{\PC}[1]{\ensuremath{\left(#1\right)}} 
\newcommand{\PX}[1]{\ensuremath{\left\lbrace#1\right\rbrace}} 
\newcommand{\BR}[1]{\ensuremath{\left\langle#1\right\vert}} 
\newcommand{\KT}[1]{\ensuremath{\left\vert#1\right\rangle}} 
\newcommand{\MD}[1]{\ensuremath{\left\vert#1\right\vert}} 

\title{Evolution of Fermion Resonance in Thick Brane}


\author{Chun-Chun Zhu$^{a}$$^{b}$$^{c}$}
\author{Qin Tan$^{a}$$^{b}$$^{c}$}
\author{Yu-Peng Zhang$^{a}$$^{b}$$^{c}$}
\author{Yu-Xiao Liu$^{a}$$^{b}$$^{c}$ \footnote{liuyx@lzu.edu.cn, corresponding author}}


\affiliation{	$^{a}$Institute of Theoretical Physics $\&$ Research Center of Gravitation, Lanzhou University, Lanzhou 730000, China\\
	$^{b}$Key Laboratory of Quantum Theory and Applications of MoE, Lanzhou University, Lanzhou 730000, China\\
	$^{c}$Lanzhou Center for Theoretical Physics $\&$ Key Laboratory of Theoretical Physics of Gansu Province, Lanzhou University, Lanzhou 730000, China}

\begin{abstract}
   {In this work, we investigate numerical evolution of massive Kaluza-Klein (KK) modes of a Dirac field on a thick brane. We deduce the Dirac equation in five-dimensional spacetime, and obtain the time-dependent evolution equation and Schr\"odinger-like equation of the extra-dimensional component. We use the Dirac KK resonances as the initial data and study the corresponding dynamics. By monitoring the decay law of the left- and right-chiral KK resonances, we compute the corresponding lifetimes and find that there could exist long-lived KK modes on the brane. Especially, for the lightest KK resonance with a large coupling parameter and a large three momentum, it will have an extremely long lifetime.}

\end{abstract}

\maketitle

\section{Introduction}

In 1914, Nordstr\"om first proposed the concept of extra dimensions~\cite{Nordstrom:1914ejq,Nordstrom:1914fn}. In the 1920s, Klein and Kaluza introduced the extra dimension theory, also known as the Klein-Kaluza (KK) theory, to unify the four-dimensional electromagnetic and gravitational interactions~\cite{kaluza:1921un,Klein:1926tv}. However, there was no noticeable development in the following decades. Until the establishment of quantum field theory, the gauge hierarchy problem (fine-tuning problem) appeared in the Standard Model. The simple description is why the Plank scale is about 16 orders of magnitude higher than the weak scale. In the 1990s, in order to solve the hierarchy problem, Arkani-Hamed, Dimopoulos, and Dvali (ADD) proposed the large extra-dimensional theory~\cite{Arkani-Hamed:1998jmv}. Then, Antoniadis, Arkani-Hamed, Dimopoulos, and Dvali first embed the brane world model into string theory~\cite{Antoniadis:1998ig}.  However, the ADD model only shifted the hierarchy problem to the scale of the extra dimension. In 1999, Randall and Sundrum (RS) proposed a warped extra-dimensional model, which successfully solved the hierarchy problem without the shortcoming of the ADD model~\cite{Randall:1999ee}.  In the same year, they  proposed another extra-dimensional model that can realize the four-dimensional Newtonian potential on the brane even though the size of extra dimension is infinite~\cite{Randall:1999vf}. The thickness of the RS brane is infinitely thin. But can there be a thick brane with infinite one or more extra dimensions? Based on this idea, the thick brane theory was developed~\cite{DeWolfe:1999cp,Gremm:1999pj,Gremm:2000dj}. That is, matter is localized on the brane by the effective potential generated by the space-time background. After that, various related extra-dimensional models were developed~\cite{Goldberger:1999uk,Gremm:1999pj,DeWolfe:1999cp,Bazeia:2008zx,Charmousis:2001hg,Arias:2002ew,Barcelo:2003wq,Bazeia:2004dh,Castillo-Felisola:2004omi,Kanno:2004nr,Barbosa-Cendejas:2005vog,Koerber:2008rx,Barbosa-Cendejas:2007cwl,Johnson:2008kc,Liu:2011wi,Chumbes:2011zt,Andrianov:2012ae,Kulaxizi:2014yxa,deSouzaDutra:2014ddw,Chakraborty:2015zxc,Karam:2018squ}.

In thick brane models, it is expected that the zero modes of various fields will be localized on the brane. The localization mechanism of these fields on the brane has been well established in different gravitational theories~\cite{Gregory:2000jc,Guerrero:2009ac,Herrera-Aguilar:2010ehj,Sousa:2012jw,Pomarol:1999ad,Oda:2000zc,Ghoroku:2001zu,Bazeia:2004dh,Dvali:1996xe,German:2012rv,Melfo:2006hh,Gremm:2000dj,Gremm:1999pj,Sousa:2014dpa,Bajc:1999mh,Liu:2008wd,Liu:2009dt,Zhang:2016ksq,Liu:2013kxz,Liang:2009zzg,Xie:2019jkq,Liu:2011zy,Li:2017dkw,Zhong:2022wlw}. Since fundamental matters consist of fermions, the nature of spin 1/2 fermions is important in braneworld models. Typically, interactions with background fields are introduced to enable the localization of either left- or right-chiral fermion zero mode on the brane~\cite{Melfo:2006hh,Liu:2013kxz,Zhang:2016ksq,Liang:2009zzg,Xie:2019jkq,Liu:2011zy,Li:2017dkw}. Usually, the massive fermion KK modes are not localized on the brane. However, some of them may be quasi-localized on the brane. Such massive modes are known as fermion resonances. It is worth noting that the resonances are fundamentally a unique category of quasi-normal modes in the context of thick brane backgrounds, known as quasi-normal resonant modes~\cite{Momennia:2018hsm,Zhou:2013dra,Barranco:2012qs,Gajic:2019qdd,Hod:2016aoe}. Previous literature has investigated the resonances of various fields in thick brane models~\cite{Liu:2009ve,Almeida:2009jc,Cruz:2013uwa,Xu:2014jda,Csaki:2000pp,Zhang:2016ksq,Sui:2020fty,Tan:2020sys,Chen:2020zzs}. But the evolution of these resonances has been rarely studied.

The detection of extra dimensions is one of the most exciting questions in physics.  Various models of extra dimensions offer distinct experimental implications. For instance, in certain thick brane models, the presence of KK gravitons can induce modifications in the Newtonian gravitational potential~\cite{Gregory:2000jc,Herrera-Aguilar:2010ehj,German:2012rv,Randall:1999vf,Barbosa-Cendejas:2007cwl,Liu:2011wi,Gregory:2000jc,Herrera-Aguilar:2010ehj,German:2012rv,Liu:2009dt,Xu:2014jda,Csaki:2000pp,Tan:2020sys,Chen:2020zzs}. Moreover, the detection of gravitational waves brings attention to the possibility of observing the short-cut effect of gravitational waves predicted in some extra-dimensional models, providing an additional way for detecting the presence of extra dimensions~\cite{Lin:2022hus,Chung:1999xg,Abdalla:2001he}.  Recently, we have numerically studied the dynamic of the scalar resonance on a thick brane and showed that the lifetime of the scalar resonance can be long enough to reach the age of the universe, which provides the possibility that the scalar resonances can be considered as a potential candidate for dark matter~\cite{Tan:2022uex}. A natural question is whether there are sufficiently long-lived resonances on a thick brane for a high dimensional Dirac field. While the neutrinos are currently one of the most favored candidates for dark matter among fermions~\cite{Dodelson:1993je}, it is worth noting that usual fermions, such as electrons and quarks, are more readily detectable due to their stronger interactions with the photon. Consequently, long-lived KK fermions, which arise in theories with extra dimensions, present an advantageous way for exploring the existence of these extra spatial dimensions. The extended lifetimes of these KK fermions facilitate their detection and  can be employed as a characteristic mode for probing the presence of extra dimensions in experimental investigations. Thus, despite not being directly associated with dark matter, these long-lived KK fermions offer a possible way of verifying the existence of extra dimensions. Therefore, in this paper, we would like to investigate dynamical evolution of fermion resonances in a thick brane model.

The remaining part of this paper is structured as follows. In Sec.~\ref{5fieldeq}, we introduce the localization mechanism of a five-dimensional Dirac fermion on a brane and derive the dynamical evolution equations of the Dirac field and the equations of motion of the left- and right-chiral fermion KK modes. In Sec.~\ref{5Dresonance}, we obtain the resonances of the five-dimensional fermion and derive their lifetimes in terms of  the full width at half maximum. In Sec~\ref{5Devolution}, we obtain the half-life through the decay of energy by numerically solving the evolution equations.  Finally, we give the discussions and conclusions in Sec.~\ref{conclusion}.

\section{Five-dimensional Dirac field}~\label{5fieldeq}

First, we review the localization of a five-dimensional fermion on a brane~\cite{Zhang:2016ksq,Liang:2009zzg,Xie:2019jkq,Liu:2011zy,Li:2017dkw}. The mechanism is implemented by introducing a coupling between the background scalar field and the fermion field. In this paper, we consider the following action of a five-dimensional Dirac fermion with the Yukawa coupling~\cite{Rubakov:1983bb,Arkani-Hamed:1999ylh,Ringeval:2001cq}
\begin{eqnarray}
	S_{\frac{1}{2}}=\int \dd^5x\sqrt{-g}~
	\Big[
	\bar{\Psi}\Gamma^M(\partial_M+\omega_M)\Psi
	+\eta\bar{\Psi}F(\phi)\Psi
	\Big],\label{action}
\end{eqnarray}
where $F(\phi)$ is function of the background scalar field $\phi$ and $\eta$ is the Yukawa coupling parameter. In  five-dimensional spacetime, Dirac field $\Psi$ is a four-component spinor. The spin connection $\omega_M$ is given by
\begin{equation}
	\omega_M=\frac{1}{4}\omega_M^{\,\,\,\,\bar{M}\bar{N}}\Gamma_{\bar{M}}\Gamma_{\bar{N}},
	\label{spin connection}
\end{equation}
where
\begin{eqnarray}
	\omega_M^{\,\,\,\,\bar{M}\bar{N}}
	&=&\frac{1}{2}E^{N\bar{M}}(\partial_ME^{\,\,\,\,\bar{N}}_N-\partial_NE^{\,\,\,\,\bar{N}}_M)\nonumber\\
	&&-\frac{1}{2}E^{N\bar{N}}(\partial_ME^{\,\,\,\,\bar{M}}_N-\partial_NE^{\,\,\,\,\bar{M}}_M)\nonumber\\
	&&-\frac{1}{2}E^{P\bar{M}}E^{Q\bar{N}}E^{\,\,\,\,\bar{R}}_M(\partial_P E_{Q\bar{R}}-\partial_Q E_{P\bar{R}}).\label{spin connection1}
\end{eqnarray}
Here, capital Latin letters $M, N, \dots=0,1,2,3,5$  label the five-dimensional spacetime indices, while letters $\bar{M}, \bar{N}, \dots=0,1,2,3,5$ label Lorentz ones. The Gamma matrices
satisfy $\{\Gamma^M,\Gamma^N\}=2g^{MN}$ and the vielbein $E^M_{\,\,\,\,\bar M}$ satisfies $E^M_{\,\,\,\,\bar{M}}E^N_{\,\,\,\,\bar{N}} \eta^{\bar{M}\bar{N}} = g^{MN}$.
\par We consider a static flat brane that satisfies four-dimensional Poincar\'e invariance on the brane, just as the scenario considered in the work of Randall and Sundrum~\cite{Randall:1999vf}. This means that the induced metric at every point of the extra dimension is a four-dimensional flat metric, and the component of the five-dimensional metric is only related to the extra dimensional coordinate $y$. The metric form that satisfies the above conditions is ~\cite{Randall:1999vf,Randall:1999ee,Halyo:1999zx,Chamblin:1999cj,Liu:2008wd,Rubakov:1983bb}
\begin{eqnarray}
	ds^2&=&g_{MN}dx^Mdx^N \nonumber\\
	&=&e^{2A(y)}\eta_{\mu\nu} \dd x^\mu \dd x^\nu+\dd y^2, \label{metric}
\end{eqnarray}
where $\e^{2A(y)}$ is the warp factor.
With the coordinate transformation $dy=\e^{A(z)} \dd z$, the above metric~(\ref{metric}) can be rewritten as
\begin{eqnarray}
	\dd s^2=\e^{2A(z)}(\eta_{\mu\nu} \dd x^\mu \dd x^\nu+\dd z^2), \label{conformal metric}
\end{eqnarray}
which is very useful in the following discussion of the equations of motion for the Dirac field. Here, Greek letter $\mu, \nu=0,1,2,3$ label the four-dimensional spacetime indices. We assume that the warp factor $\e^{2A(y)}$ and the scalar field $\phi(y)$ are only functions of the extra dimension coordinate $y$. From the conformal flat metric~(\ref{conformal metric}),
the nonvanishing components of the spin connection~(\ref{spin connection}) are  $\omega_\mu=\frac{1}{2}\partial_z A\gamma_\mu\gamma_5$. The five-dimensional Dirac equation can be obtained by varying the action as follows
\begin{equation}
	\Big[\gamma^{\mu}\partial_\mu
	+\gamma^5(\partial_z+2\partial_z A)+\eta F\Big]\Psi=0.
	\label{newcouplingmotion}
\end{equation}
Then we introduce the following chiral decomposition of the Dirac field:
\begin{eqnarray}
	\Psi(x^i,t,z)&=&\e^{-2A}\Psi'(x^i,t,z)\nonumber\\
	&=&\e^{-2A}\sum_{n}\Big[\psi_{\text{L}n}(x^i)F_{\text{L}n}(t,z)\nonumber\\
	& &+\psi_{\text{R}n}(x^i)F_{\text{R}n}(t,z)\Big],
	\label{decomposition}
\end{eqnarray}
where $i=1,2,3$ label three-dimensional space indices, $\psi_{\text{L}n}=-\gamma^5\psi_{\text{L}n}$ and $\psi_{\text{R}n}=\gamma^5\psi_{\text{R}n}$ are left- and right-chiral three-dimensional space components of the Dirac field, respectively. Substituting the chiral decomposition~(\ref{decomposition}) into Eq.~(\ref{newcouplingmotion}), we can rewrite Eq.~(\ref{newcouplingmotion}) as
\begin{equation}
	\Big[\gamma^{\mu}\partial_\mu
	+\gamma^5\partial_z+\eta F\Big]\Psi'=0,
	\label{newcouplingmotion2}
\end{equation}
or as the chiral form
\begin{eqnarray}
	\begin{array}{c}
	\ii(\partial_t+\sigma^i\partial_i)F_{\text{R}n}\psi_{\text{R}n} = (\partial_z+\eta F)F_{\text{L}n}\psi_{\text{L}n},\quad \\
	\ii(\partial_t-\sigma^i\partial_i)F_{\text{L}n}\psi_{\text{L}n} = (-\partial_z+\eta F)F_{\text{R}n}\psi_{\text{R}n},
	\end{array}
	\label{KGEQ}
\end{eqnarray}
where $\sigma^i$ are Pauli matrixs. The above equations~(\ref{KGEQ}) can be rewritten as
\begin{eqnarray}
	\begin{array}{c}
		\left[\partial_t^2-\partial_i^2-\partial_z^2+V_\text{L}(z)\right]F_{\text{L}n}\psi_{\text{L}n} = 0, \\
		\left[\partial_t^2-\partial_i^2-\partial_z^2+V_\text{R}(z)\right]F_{\text{R}n}\psi_{\text{R}n} = 0.
	\end{array}
	\label{secondordereq}
\end{eqnarray}
We consider free modes on the brane, i.e., $\psi_{\text{L}n,\text{R}n}(x^i)=\text{e}^{-\ii a_{ni}x^i}\chi_{\text{L}n,\text{R}n}$, where $\chi_{\text{L}n,\text{R}n}$ are four-dimensional spinors independent of the coordinate and $a_{ni}$ corresponds to the spatial momentum of the spinors on the brane. Then we can obtain the following evolution equations:
\begin{eqnarray}
	\begin{array}{c}
		\left[\partial_t^2-\partial_z^2+V_\text{L}(z)\right]F_{\text{L}n} = -a^2_n F_{\text{L}n}, \\
		{\left[\partial_t^2-\partial_z^2+V_\text{R}(z)\right]F_{\text{R}n} = -a^2_n F_{\text{R}n}},
	\end{array}
	\label{schrodingerlikeequationr}
\end{eqnarray}
where $a_n = \sqrt{a_{ni}a_{n}^{i}}$ and the effective potentials $V_\text{L,R}$ are~\cite{Liang:2009zzg,Koley:2004at,Ringeval:2001cq,Liu:2008wd,Liu:2007ku,Liu:2008pi}
\begin{equation}
	V_\text{L,R}(z)=(\eta e^AF)^2\pm\partial_z(\eta e^AF).
	\label{potentialyukawa}
\end{equation}
We further decompose $F_{\text{L}n}$ and $F_{\text{R}n}$ as
\begin{equation}
	F_{\text{L}n,\text{R}n}(t,z)=\text{e}^{\ii\omega t}f_{\text{L}n,\text{R}n}(z).
	\label{wavefunctiondecomposition}
\end{equation}
Substituting the above decompositions into Eq.~(\ref{schrodingerlikeequationr}), we finally derive the Schr\"odinger-like equations
\begin{eqnarray}
	\begin{array}{c}
		[-\partial_z^2+V_\text{L}(z)]f_{Ln} = m^2_nf_{\text{L}n}, \\
		{[-\partial_z^2+V_\text{R}(z)]f_{Rn} = m^2_nf_{\text{R}n}},
	\end{array}
	\label{schrodingerlikeequationl}
\end{eqnarray}
where $m_{n}$ is the mass of the corresponding Dirac KK modes.

\section{Fermion localization and resonance}~\label{5Dresonance}

In this section, we investigate the resonances of a bulk fermion and their evolution in the thick brane model. The action of the thick brane is~\cite{Bazeia:2006ef,Kobayashi:2001jd,Afonso:2006gi,Gremm:2000dj,Liang:2009zzg,DeWolfe:1999cp,Gremm:1999pj}
\begin{eqnarray}
	S=\int{\dd^5x\sqrt{-g}~\left[\frac{M_5^3}{4}R-\frac{1}{2}\partial_M\phi\partial^M\phi-V(\phi)\right]}.
\end{eqnarray}
We set the fundamental mass scale $M_{5}=1$ for convenience.  A flat brane solution was studied in Refs. \cite{Afonso:2006gi,Gremm:2000dj,Li:2017dkw,Liang:2009zzg}
\begin{eqnarray}
	V(\phi)&=&\frac{9}{8}k^2b^2\cosh^2(b\phi)-3k^2\sinh^2(b\phi),\label{phi}\\
	\phi(y)&=&\frac{1}{b}\text{arcsinh}\big[\tan(\frac{3}{2}kb^2y)\big],\label{potentialphi}\\
	A(y)&=&-\frac{2}{3b^2}\ln\big[\sec(\frac{3}{2}kb^2y)\big],\label{warpfactorscalar}
\end{eqnarray}
where $k, b$ are real parameters. In this paper, we choose $b=\sqrt{2/3}$.  The conformal coordinate is
\begin{eqnarray}
	z=\int_0^y \text{e}^{-A(y)}dy=\frac{1}{k}\text{arcsinh}\big[\tan(k y)\big].\label{relationyz}
\end{eqnarray}
Substituting the relation~(\ref{relationyz}) into Eqs.~(\ref{potentialphi}) and~(\ref{warpfactorscalar}), we can get the form of the scalar field and the warp factor under the conformal coordinate $z$:
\begin{eqnarray}
	\phi(z)&=&\sqrt{\frac{3}{2}}kz,\label{potentialphiz}\\
	A(z)&=& \ln\big[\text{sech}(k z)\big].  \label{warpfactorscalarz}
\end{eqnarray}

\begin{figure}[htbp]
	\centering
	\includegraphics[width=0.235\textwidth]{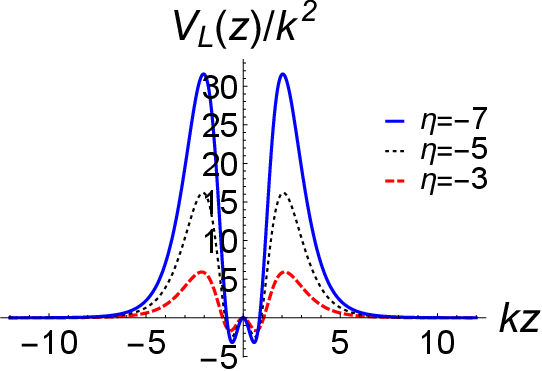}
	\includegraphics[width=0.235\textwidth]{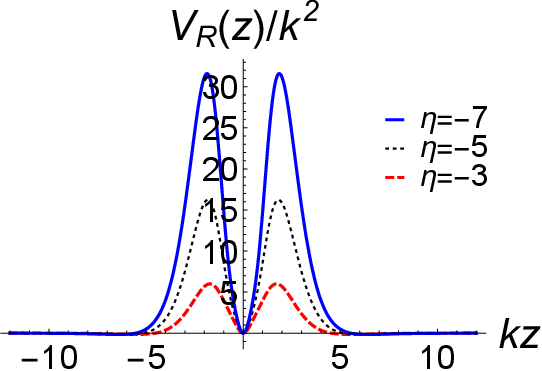}
	\vskip -4mm \caption{The shapes of the effective potentials (\ref{potential_q}) for the left- and right-chiral fermions.  }
	\label{FigPotenialsYukLambda}
\end{figure}

\begin{figure}[!htbp]
	\centering
	\subfigure[~effective potential, $\eta=-3$]{
		\centering
		\includegraphics[width=0.22\textwidth]{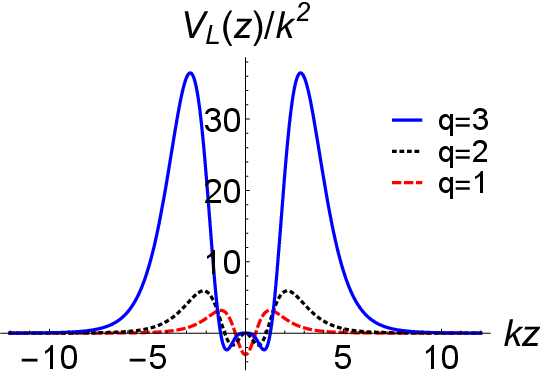}
	}
	\subfigure[~$q=1, \eta=-3$]{
		\centering
		\includegraphics[width=0.22\textwidth]{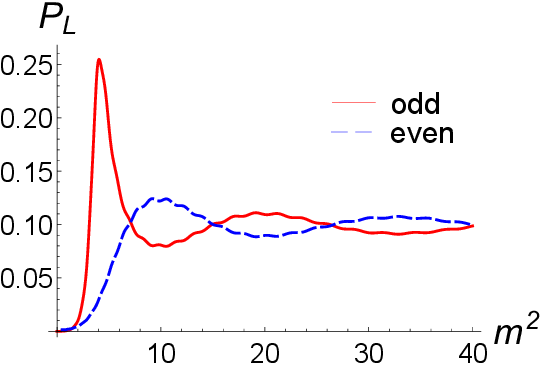}
	}
	\subfigure[~$q=2, \eta=-3$]{
		\centering
		\includegraphics[width=0.22\textwidth]{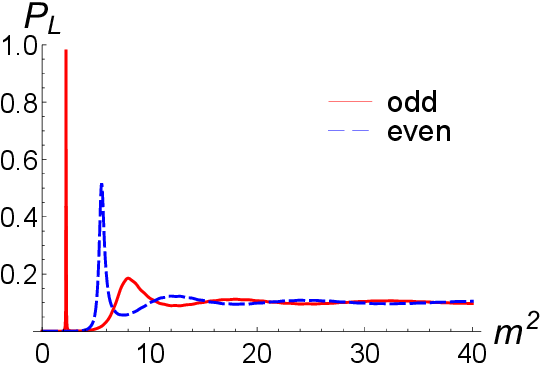}
	}
	\subfigure[~$q=3, \eta=-3$]{
		\centering
		\includegraphics[width=0.22\textwidth]{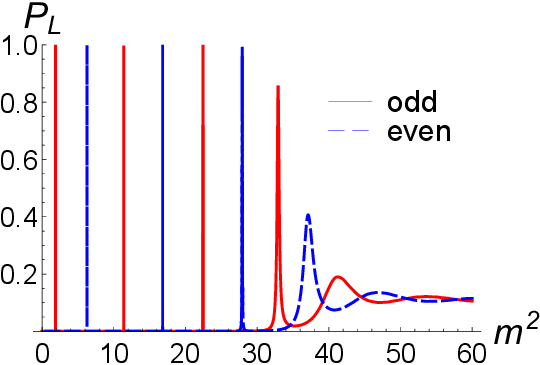}
	}
	
	\vskip -4mm \caption{Plots of the effective potentials and the relative probabilities for the left-chiral fermion for different values of $q$.}
	\label{Figprobabilityq}
\end{figure}

To study the localization of a five-dimensional fermion on the thick brane, we consider the Yukawa coupling $\eta\bar{\Psi}F(\phi)\Psi$ between the fermion $\Psi$ and the background scalar field $\phi$. The coupling function is chosen as $F(\phi)=k~\text{arcsinh}^{2q-1}(b\phi)$, where the structure parameter $q$ is a positive integer~\cite{Zhang:2016ksq}. By substituting the specific form of $F(\phi)$ into Eq.~(\ref{potentialyukawa}), we can obtain the expression of the effective potential:
\begin{eqnarray}
	V_{\text{L,R}}(z)&=&\pm\eta~k^2~\text{sech}\left(k z \right)\text{arcsinh}^{2q-2}
	\left(k z \right)\times \nonumber\\
	&&\Bigg(\frac{(2q-1)} {\sqrt{1+k^2 z^2}  
	\pm\eta~\text{arcsinh}^{2 q}\left(k z\right)\times} \nonumber\\
	&&\text{sech}\left(k z \right)-\text{arcsinh}\left(k z\right) \tanh\left(k z\right)
	\Bigg).   \label{potential_q}
\end{eqnarray}
We plot the effective potentials with different values of the coupling parameter $\eta$ in Fig.~\ref{FigPotenialsYukLambda}. Note that we always have $V_{\text{L,R}}(0)=0$.  It can be seen that the stronger the coupling, the deeper the (quasi) potential wells. In addition to the coupling parameter $\eta$, there are also two parameters $k$ and $q$ that affect the potential functions. We also give the plots of the potentials with different values of $q$ in Fig.~\ref{Figprobabilityq}. Obviously, the similar behaviors induced by the coupling parameter $\eta$ are also found when changing the parameter $q$, i.e., the depth of the quasi potential well increases with $q$. Note that the parameter $k$  only rescales the eigenvalues of Schr\"odinger-like
equations~(\ref{schrodingerlikeequationl}). For simplicity, here we do not consider the impact of the change in $k$ on the potential function and take $k=1$.

The solutions of the left and right-chiral zero modes are
\begin{equation}
	f_{\text{L}0,\text{R}0}\propto
	\e^{ \pm\eta\int F(\phi) {\dd z}}=\e^{\pm\eta\int \text{arcsinh}^{2q-1}(\phi) {\dd z}}\stackrel{z\rightarrow\infty}{\longrightarrow}\text{e}^{\pm \eta z(\ln z)^{2q-1}}. \label{newzero mode}
\end{equation}
It can be seen that, when the coupling $\eta$ is negative, only the left-chiral zero mode satisfies
\begin{eqnarray}
	\int_{-\infty}^\infty{|f_{\text{L}0}|^2\dd z}<\infty.
	\label{zero modeyukawa}
\end{eqnarray}
Thus, only the left-chiral zero mode can be localized on the brane for a negative coupling~\cite{Zhang:2016ksq,Liang:2009zzg,Xie:2019jkq,Liu:2011zy,Li:2017dkw}. Conversely, only the right-chiral zero mode can be localized on the brane for a positive coupling.

Next, we investigate the characteristics of fermion resonances for this braneworld model. Apart from the zero mode, there exist some massive KK modes that are quasi-localized on the brane, which are called as resonances. In 2009, Almeida et al. defined resonances by using the peak value of the  square of a normalized wave function at a fixed point in the well of the effective potential~\cite{Almeida:2009jc}. Subsequently, the relative probability method was introduced to look for all resonances~\cite{Liu:2009ve}. In 2011, the transfer matrix method was used to obtain resonances~\cite{Landim:2011ki}. A resonance is a massive KK mode quasi-localized within an effective (quasi) potential well, i.e., quasi-localized on the brane. For a resonance, its transmitted and reflected oscillatory modes are in phase in the effective (quasi) potential well. A resonant state usually has a larger amplitude within the brane. Thus, one can employ the probability ratio of a massive KK mode inside the brane $(-z_b,z_b)$ and inside a wider range of extra dimension $(-z_\text{max},z_\text{max})$ to quantify the reflectivity in a relatively straightforward manner, without the need for a transfer matrix approach. Specifically, we consider the range of extra dimension to be $n$ times of the brane, i.e., $z_\text{max}=n z_b$. For a plane wave mode, the probability ratio is $1/n$. If the probability ratio exceeds $1/n$, the reflection will exceed the transmittance and the massive mode might be a resonance. In this paper, we use the relative probability method to find all resonances. The relative probability is defined as~\cite{Liu:2009ve}
\begin{eqnarray}
	P_\text{L,R}(m^2)=\frac{\int_{-z_b}^{z_b}|f_{\text{L}n,\text{R}n}(z)|^2 \dd z}
	{\int_{-z_\text{max}}^{z_\text{max}}|f_{\text{L}n,\text{R}n}(z)|^2 \dd z},
	\label{zbmin}
\end{eqnarray}
where $z_\text{max}=n z_{b}$ and $z_b$ is approximately the width of the brane.
Each local maximum of the relative probability with a full width at half maximum corresponds to a resonance.
We focus on the dynamic behavior of the resonances in the brane, so $z_b$ is taken as the coordinate value corresponding to the maximum of the potential. If the range of extra dimension is too small, the relative probability error will be large. For a not too small $z_\text{max}$, e.g., $z_\text{max} \geq 10 z_b$, its value has no effect on the resonance spectrum. Here, we take $n=10$.  Since the effective potential is symmetric, we can take the following boundary conditions to obtain the massive KK modes numerically:
\begin{eqnarray}
	\label{incondition}
	f_{\text{L}n,\text{R}n}(0)&=&0, ~f'_{\text{L}n,\text{R}n}(0)=1,~\text{odd KK modes},\\
	f_{\text{L}n,\text{R}n}(0)&=&1, ~f'_{\text{L}n,\text{R}n}(0)=0,~\text{even KK modes}.
\end{eqnarray}

\begin{figure}[!h]
	\centering
	\subfigure[~$\eta=-3$, left-chiral]{
		\centering
		\includegraphics[width=0.22\textwidth]{eta3P.eps}
	}
	\subfigure[~$\eta=-3$, right-chiral]{
		\centering
		\includegraphics[width=0.22\textwidth]{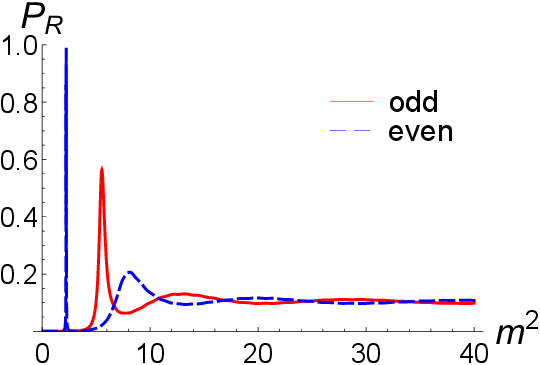}
	}
	\subfigure[~$\eta=-5$, left-chiral]{
		\centering
		\includegraphics[width=0.22\textwidth]{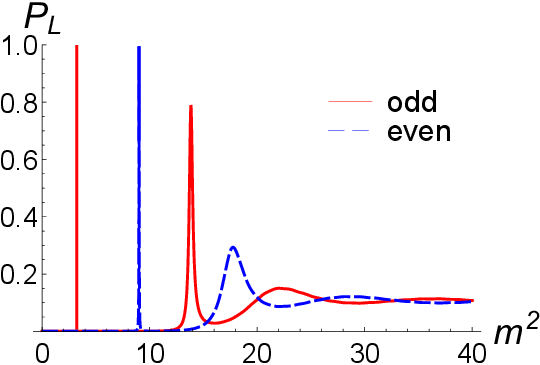}
	}
	\subfigure[~$\eta=-5$, right-chiral]{
		\centering
		\includegraphics[width=0.22\textwidth]{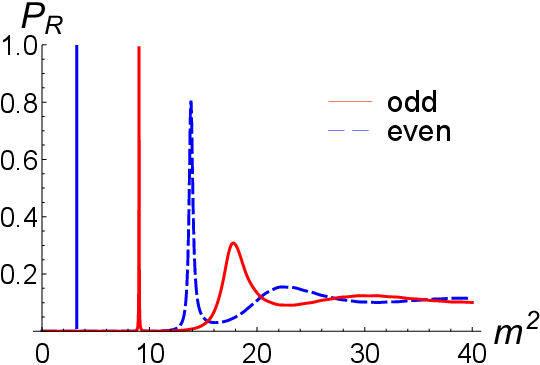}
	}
	\subfigure[~$\eta=-7$, left-chiral]{
		\centering
		\includegraphics[width=0.22\textwidth]{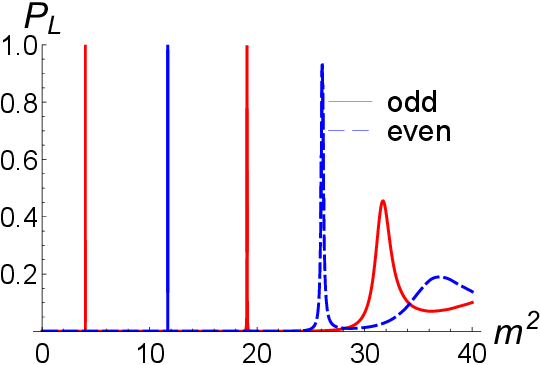}
	}
	\subfigure[~$\eta=-7$, right-chiral]{
		\centering
		\includegraphics[width=0.22\textwidth]{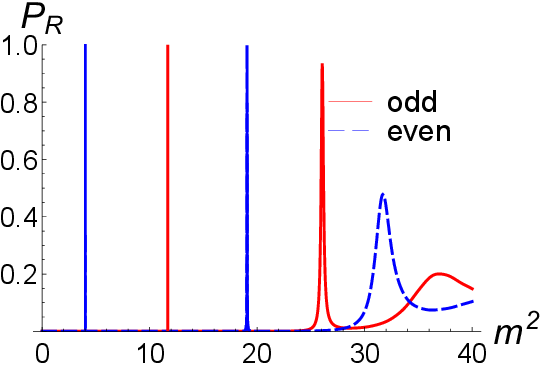}
	}
	
	\vskip -4mm \caption{Plots of the relative probabilities of the left- and right-chiral fermion KK modes for different values of $\eta$ and fixed value $q=2$. The red lines are odd-parity modes and the blue dashed lines are even-parity modes.}
	\label{Figprobability7}
\end{figure}

\begin{figure}[h]
	\centering
	\subfigure[~The first left-chiral fermion resonance with $m_1^2=4.0289$.]{
		\centering
		\includegraphics[width=0.22\textwidth]{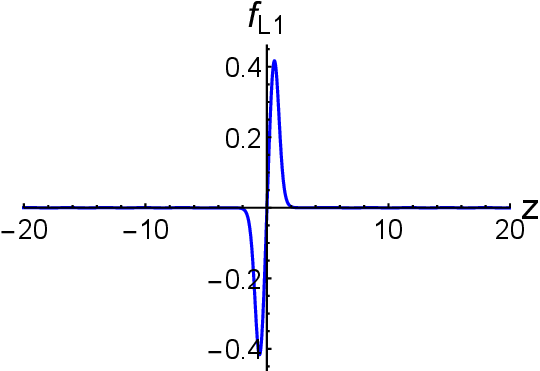}
	}
	\subfigure[~The first right-chiral fermion resonance with $m_1^2=4.0289$.]{
		\centering
		\includegraphics[width=0.22\textwidth]{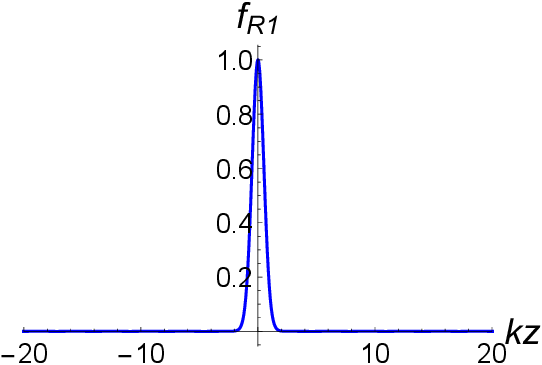}
	}
	\subfigure[~The second left-chiral fermion resonance with $m_2^2=11.6926$.]{
		\centering
		\includegraphics[width=0.22\textwidth]{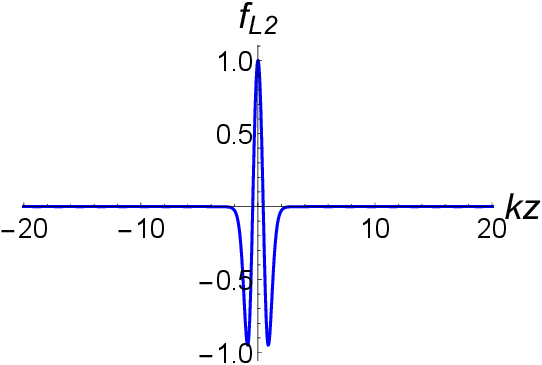}
	}
	\subfigure[~The second right-chiral fermion resonance with $m_2^2=11.6926$.]{
		\centering
		\includegraphics[width=0.22\textwidth]{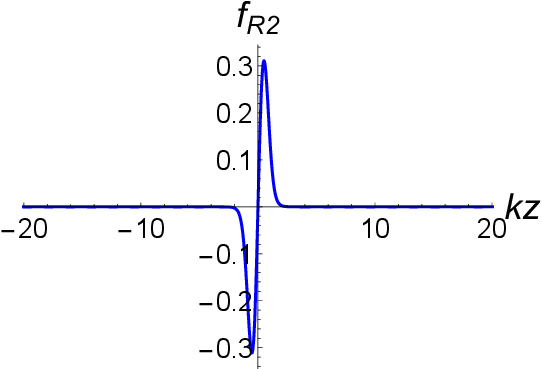}
	}
	\subfigure[~The left-chiral fermion non-resonance with $m^2=13$. ]{
		\centering
		\includegraphics[width=0.22\textwidth]{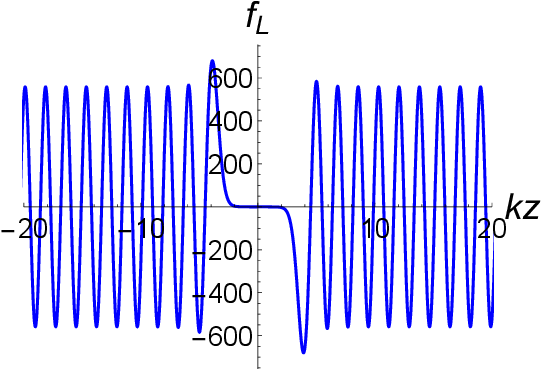}
	}
	\subfigure[~The right-chiral fermion non-resonance with $m^2=13$.]{
		\centering
		\includegraphics[width=0.22\textwidth]{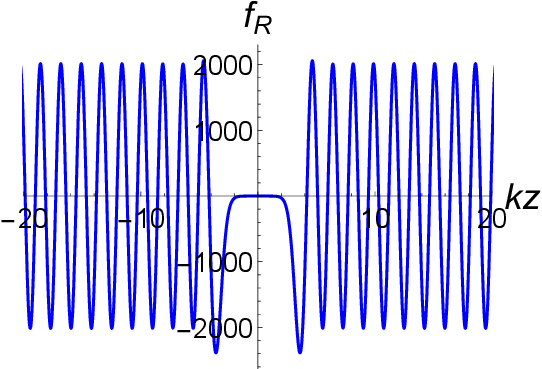}
	}
	
	\vskip -4mm \caption{Plots of the wave functions of different chiral resonances.  }
	\label{Figwavefuction}
\end{figure}

For a KK mode with the mass $m_n$, if its relative probability $P(m^2)$ has just a peak at $m = m_n$ and this peak has a full width at half maximum, we say that there is a resonance with the mass $m_n$. The relative probabilities of a left-chiral fermion's massive mode for different $q$ are shown in Fig.~\ref{Figprobabilityq}. It shows that the depth and width of the quasi well significantly increase with the parameter $q$. It means that the number and the relative probabilities of resonances also increase greatly with $q$. The relative probabilities of the massive KK modes of the left- and right-chiral fermions  for different $\eta$ are shown in Fig.~\ref{Figprobability7}. It is evident that the peak value and the number of resonances increase with $|\eta|$. Since the resonance cannot be localized on the brane, it can only stay on the brane for a finite time. Here, we use the full width at half maximum to roughly estimate the lifetime of the resonances. Usually, the first resonance has the longest lifetime. It should be noted that, in other thick brane models, the first resonance may not be  the longest-lived mode~\cite{Xie:2019jkq,Farokhtabar:2016fhm}. In our model, we can see that the relative probability of the resonance decreases with the mass $m_n$, correspondingly, the  lifetime of the resonance also decreases with the mass $m_n$. We can also see that the left- and right-chiral fermions share the same resonant spectrum. The reason is that the finite effective potentials (\ref{potentialyukawa}) of the left- and right-chiral fermions are supersymmetric partner potentials, and there are finite in the whole region. So they have exactly the same mass spectrum except for the zero mode, which was also discussed in Ref.~\cite{Zhang:2016ksq,Liang:2009zzg,Liu:2008wd,Liu:2007ku,Liu:2008pi}. Therefore, even though the right-chiral fermion lacks a zero mode and cannot be fully localized on the brane for negative coupling parameter $\eta$, some long-lived modes can be quasi-localized on the brane, which enables the possibility of detecting right-chiral fermions. The wave functions of the resonances and nonresonances for the case of $\eta=-7$ and $q=2$ are shown in Fig.~\ref{Figwavefuction}. It can be seen that, for a resonance, the amplitude of its wave function inside the potential well is much larger than that outside the well. The opposite is true for nonresonances. So far, we have introduced the relative probability method to derive the resonances. Next, we will focus on the evolution of them.

\section{Resonance evolution}~\label{5Devolution}

In this section, we investigate the evolution behavior of fermion resonances. Under the stationary assumption, one can obtain such resonances in terms of the relative probability. Since these resonances are not localized on the brane, they cannot exist stably on it. Therefore, one can take them as the initial data and  numerically evolve them by using Eq.~(\ref{schrodingerlikeequationr}) with the radiation boundary condition~\cite{Megevand:2007uy}.

Note that, a resonance can not be localized on the brane and the corresponding total energy is divergent. Such divergent total energy means that one can not use it to grasp the dynamical properties of the resonance. To overcome this disadvantage, we define an effective energy of the resonance in a finite range $(-z_b,z_b)$ in terms of the energy-momentum tensor. The specific form of the energy-momentum tensor $T_{MN}$ in the case of the action~(\ref{action}) in five-dimensional spacetime is
\begin{eqnarray}
	T_{MN}=\frac{1}{2}(\bar{\Psi}\Gamma_M (\partial_N+\omega_N)\Psi
	+\bar{\Psi}\Gamma_N (\partial_M+\omega_M)\Psi)\nonumber\\
	+g_{MN}\left(\bar{\Psi}  \Gamma^K(\partial_K+\omega_K) \Psi
	+\eta\bar{\Psi}F(\phi)\Psi\right).
	\label{energy-momentum}
\end{eqnarray}
Combining the energy-momentum tensor $T_{MN}$ and the time-like killing vector $k^N$, one can derive the conserved current as~\cite{Pavlidou:2000cs}
\begin{equation}
	J_{M}=T_{MN}k^{N}.
	\label{current}
\end{equation}
Then we can define the corresponding energy:~\cite{Pavlidou:2000cs}
\begin{equation}
	E(t)=\int J^0 \sqrt{-g} \dd^3x\dd z.
	\label{abstract energy}
\end{equation}
Substituting Eq.~(\ref{energy-momentum}) into the above equation, we can get
\begin{eqnarray}
	E(t) =\int \psi_{\text{L}}^{*} \psi_\text{L} \dd^3x \int_{-z_{b}}^{z_{b}} (2\ii F_{\text{L}}^{*}\partial_t F_{\text{L}}-a_{n}F_{\text{L}}^{*} F_{\text{L}}) \text{e}^{A} \dd z\nonumber\\
	+\int \psi_{\text{R}}^{*} \psi_{\text{R}} \dd^3x \int_{-z_{b}}^{z_{b}} (2\ii F_{\text{R}}^{*}\partial_t F_{\text{R}}+a_{n}F_{\text{R}}^{*} F_{\text{R}}) \text{e}^{A}\dd z.
	\label{specific energy}
\end{eqnarray}

In this paper, we focus on the evolution of fermion resonances along the extra dimension. Therefore, we only calculate the evolution of the extra dimension profile for the left- and right-chiral fermion KK modes separately. So the conserved energy simplifies to
\begin{equation}
	E(t)=\mathcal{B} E_e(t)=\mathcal{B} \int_{-z_{b}}^{z_{b}} \rho(t,z)\text{e}^{A}\dd z,
\end{equation}
where
\begin{eqnarray}
	\mathcal{B} &=& \int \psi_{\text{L,R}}^{*} \psi_\text{L,R} \dd^3x, \\
	E_e(t) &=& \int_{-z_{b}}^{z_{b}} \rho_{\text{L,R}}(t,z)\text{e}^{A}\dd z, \\
	\rho_{\text{L,R}}(t,z) &=& 2\ii F_{\text{L,R}}^{*}\partial_t F_{\text{L,R}}\mp a_{n}F_{\text{L,R}}^{*} F_{\text{L,R}}.
	\label{feimion enenergy compsosition}
\end{eqnarray}

\begin{figure*}[htbp]
	\centering
	\subfigure[~$\eta=-3$, left-chiral]{
		
		\includegraphics[width=0.32\textwidth]{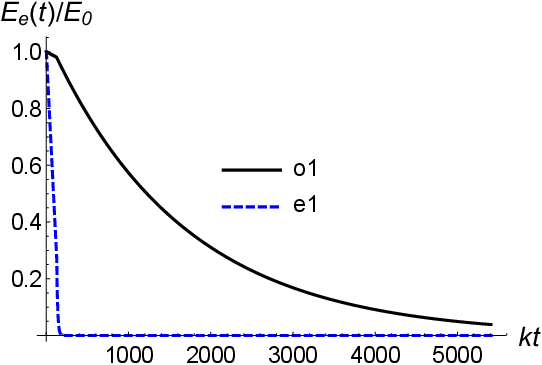}
	}
	\subfigure[~$\eta=-5$, left-chiral]{
		
		\includegraphics[width=0.32\textwidth]{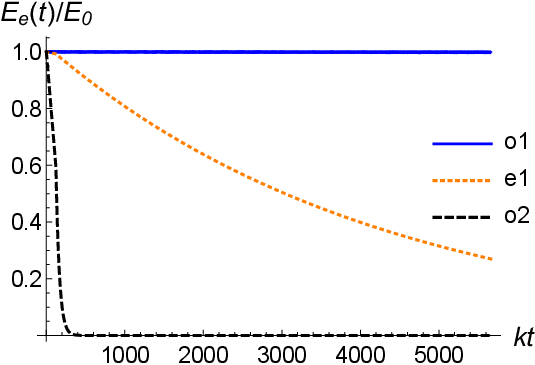}
	}
	\subfigure[~$\eta=-7$, left-chiral]{
		
		\includegraphics[width=0.32\textwidth]{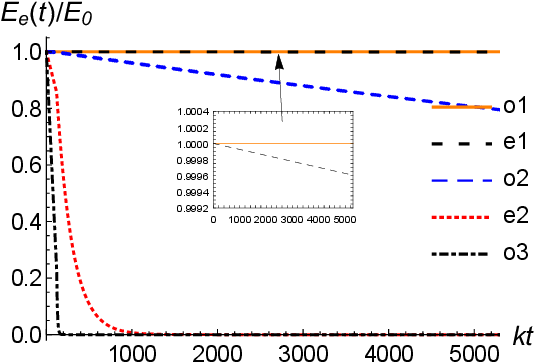}
	}
	\subfigure[~$\eta=-7$, right-chiral]{
		
		\includegraphics[width=0.32\textwidth]{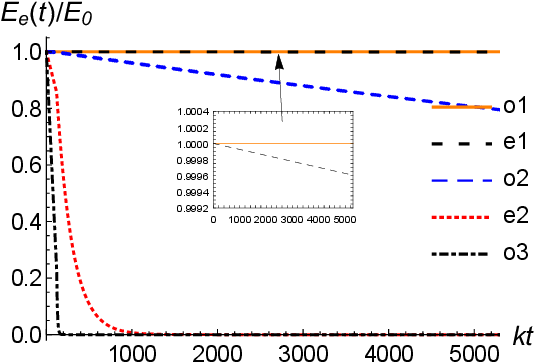}
	}
	\vskip -4mm \caption{Plots of the energy as the function of time with different $\eta$ and chirality. Parameter $q$ set to 2. Here o$j$ and e$j$ represent the $j$-th odd-parity and even-parity resonances, respectively.}
	\label{Figenenrgyevolution}
\end{figure*}

First, we consider the case of $a_{n}=0$. The evolution of energy $E_e(t)$ for each KK fermion is shown in Fig.~\ref{Figenenrgyevolution}. It can be seen that the first resonance with the smallest mass has the smallest decay rate. The decay rate of the resonance decreases with the parameter $|\eta|$. We can also fit the decay of the energy $E_e$ with the exponential form as follows
\begin{equation}
	E_e(t)=E_0 \text{e}^{-st},
\end{equation}
where $s$ is the decay parameter. We define the lifetime $\tau$ by $E_e(\tau)/E_e(0)=1/2$. Thus, the relation between the lifetime and the decay parameter is $\tau = \frac{\ln 2}{s}$. Note that the term ``decay'' solely refers to the energy loss of the fermion resonances on the brane over time. It does not involve any decay channel and decay product in particle physics. 

\begin{figure}[htbp]
	\centering
	\subfigure[~first resonance, left-chiral]{	
		\includegraphics[width=0.22\textwidth]{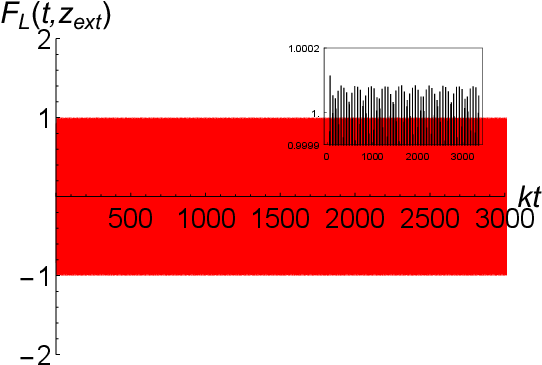}
	}
	\subfigure[~first resonance, right-chiral]{
		\includegraphics[width=0.22\textwidth]{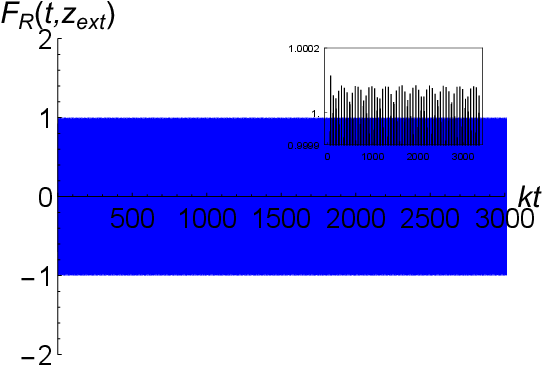}
	}
	\subfigure[~second resonance, left-chiral]{
		\includegraphics[width=0.22\textwidth]{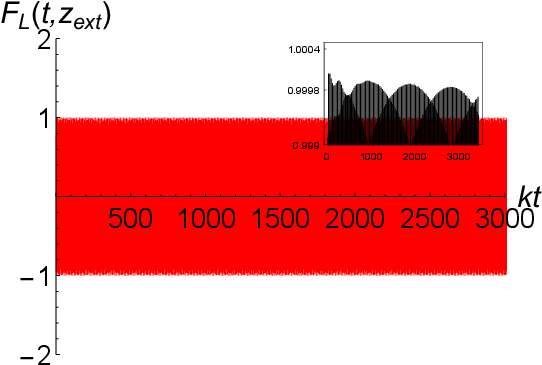}
	}
	\subfigure[~second resonance, right-chiral]{
		\includegraphics[width=0.22\textwidth]{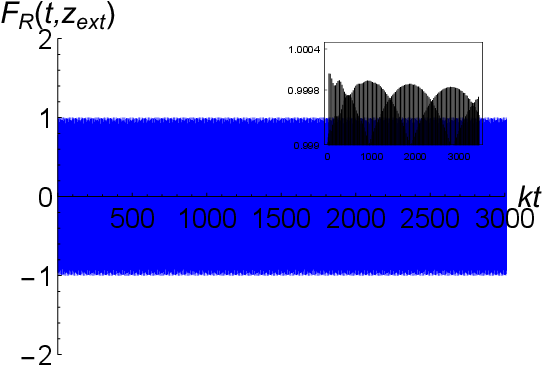}
	}
	\subfigure[~third resonance, left-chiral]{
		\includegraphics[width=0.22\textwidth]{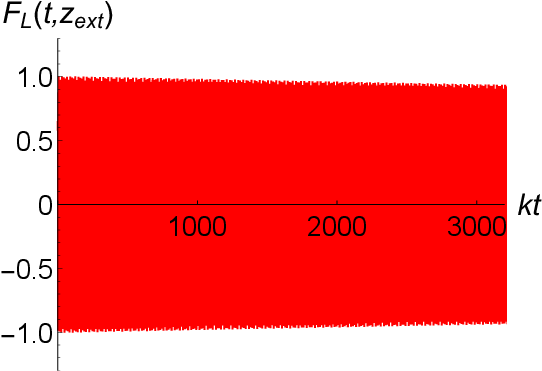}
	}
	\subfigure[~third resonance, right-chiral]{
		\includegraphics[width=0.22\textwidth]{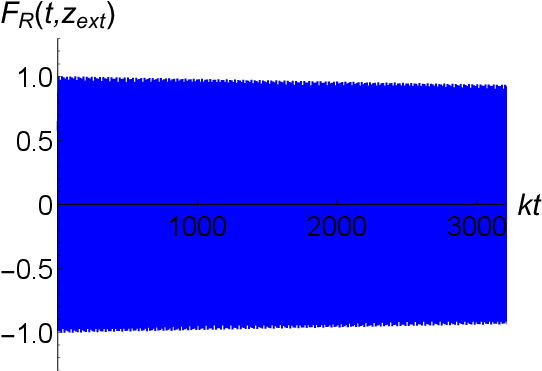}
	}
	\vskip -4mm \caption{ Time evolution of the amplitudes for the left- or right-chiral Dirac resonances at $kz_\text{ext}=2$. The parameters are set to $\eta=-7$ and $q=2$}.
	\label{Fixzevolution}
\end{figure}

\begin{figure}[htbp]
	\centering
	\includegraphics[width=0.32\textwidth]{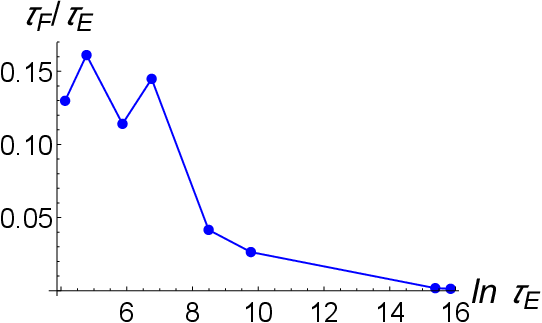}
	\vskip -4mm \caption{Plot of the relation between $\tau_F/\tau_E$ and $\ln \tau_E$. $\tau_F$ and $\tau_E$ are the lifetimes calculated by the full width at half maximum and evolution, respectively.}
	\label{figlifetime}
\end{figure}

To investigate the dynamical behavior of a Dirac resonance, we can extract a time series for the amplitude of the resonance at a fixed point $z_\text{ext}$. The parameters are set to $\eta=-7$ and $q=2$. We consider that there is no incident wave at infinity and the outgoing wave has no reflection behavior at boundaries at infinity. Therefore, we take the radiation boundary conditions at infinity on both sides: $\pd_t F_{\text{L},\text{R}}=\pm\pd_{z}F_{\text{L},\text{R}}$ for $z\rightarrow\pm\infty$. The results are shown in Fig.~\ref{Fixzevolution}, which reveals that the decay of the first resonance is slowest. This phenomenon can be attributed to the fact that the mode with higher relative probability has smaller amplitude outside the quasi-well than that inside the quasi-well. The reason is that the frequency of a resonance has a minimum transmittance. The greater the relative probability, the lower the transmittance and the slower the internal amplitude attenuation. For the boundary condition with outgoing waves on both sides, the rate of outflow of the reduced energy $E_{e}(t)/E_0$ depends on both the difference between the internal and external amplitudes and the speed of outward movement.  Thus, the greater the relative probability, the greater the difference between the internal and external amplitudes, and the slower the internal reduced energy decay. We present the results in Tab.~\ref{TableSpectraYukawa1}. It can also be seen that the lifetime calculated by using the full width at half maximum is much smaller than that obtained by evolution, and the greater the lifetime, the greater the error between two ways, which can be also seen in Fig.~\ref{figlifetime}. Therefore, the full width at half maximum can only be used to calculate the variation trend of the lifetime of the resonance very roughly, and can not get the accurate lifetime value of a resonance.

\begin{figure}[htbp]
	\centering	
	\subfigure[~energy evolution]{
		\centering
		\includegraphics[width=0.26\textwidth]{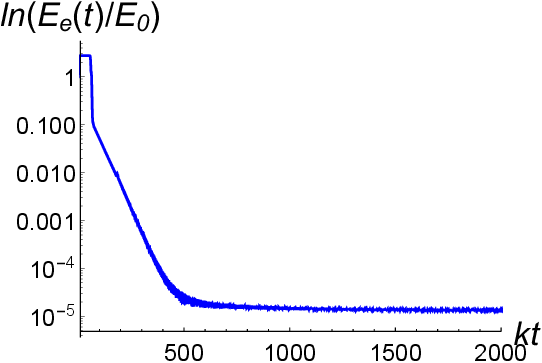}
	}
	\subfigure[~wave function evolution]{
		\centering
		\includegraphics[width=0.26\textwidth]{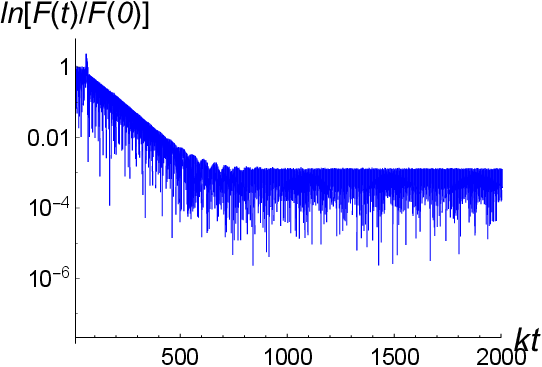}
	}	
	\vskip -4mm \caption{Time evolution of the reduced energy and the amplitude at $kz_\text{ext}=2$ for the left-chiral Dirac KK modes with $m^2$=13, $\eta=-7$ and $q=2$.}
	\label{evolutionwithm13}
\end{figure}

\begin{figure}[htbp]
	\centering
	\subfigure[~first resonance]{
		\centering
		\includegraphics[width=0.29\textwidth]{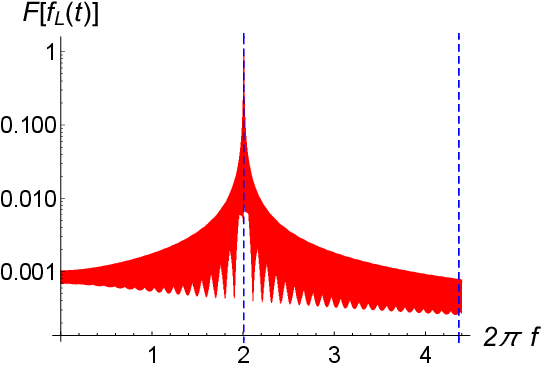}
	}
	\subfigure[~nonresonance, $m^2=13$]{
		\centering
		\includegraphics[width=0.29\textwidth]{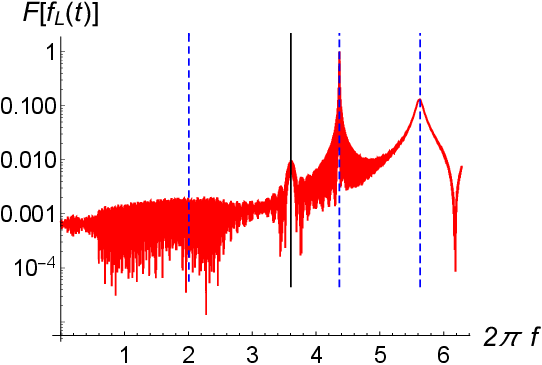}
	}
	\vskip -4mm \caption{The Fourier transform spectrum of the amplitude  at $kz_\text{ext}=2$.
		The parameters are set to $\eta=-7$ and $q=2$. The dashed blue lines correspond to the frequencies of the three odd parity resonances and the black line corresponds to the initial frequency.}
	\label{Fourierdecomposition}
\end{figure}

On the other hand, we can also treat a nonresonance as the initial data to perform its evolution. Figure~\ref{evolutionwithm13} shows the evolution of the energy and the wavefunction for the nonresonance. As expected, the amplitude and the energy of the nonresonance quickly decay at an early stage, but later they decay like those of resonances. To gain a better understanding of this phenomenon, we perform the discrete Fourier transform and present the results in Fig.~\ref{Fourierdecomposition}. The expression for the discrete Fourier transform is
\begin{eqnarray}
	F\left[ f_\text{L}(f,z_\text{ext})\right] =|A\sum_{p}f_\text{L}(t_p,z_\text{ext})~\e^{-2\pi\ii  ft_p}|,\label{discrete Fourier transform}
\end{eqnarray}
where $A=1/\text{max}(	F\left[ f_\text{L}(f,z_\text{ext})\right] )$ is a normalized constant, and $z_\text{ext}$ is a fixed point.

It is obvious that for the Fourier transform spectrum of the resonance, there is only one peak, which corresponds to the KK mass of the resonance. However, for the spectrum of the nonresonance, there are several peaks. Each peak corresponds to a resonance. This indicates that the nonresonance can evolve into a superposition of a series of resonances at late time.

In addition, we investigate the influence of the parameter $\eta$ on the lifetime of the Dirac resonances. Since the energy decay rate of the first resonance state becomes extremely small when $\eta$ is large, the accuracy error of calculation may be relatively large. Therefore, we choose the lifetime of the second resonance state here to study the relationship between its lifetime and the parameter $\eta$.  The results are shown in Tab.~\ref{TableSpectraYukawa2}. It can be seen that the lifetime of the resonance increases with $|\eta|$. This relationship can be attributed to the fact that the effective potential's height increases with $|\eta|$, thereby elevating the relative probabilities of the resonances. Since the parameter $\eta$ is the coupling strength of the background scalar field and the fermion, we can conclude that the stronger the coupling strength, the longer the lifetime of the fermion resonance on the brane.

In the previous part, we only consider the case of $a_{ni}=0$, i.e, the momentum on the brane is zero. Next we consider the case of nonvanishing $a_{ni}$.  According to Eq.~(\ref{schrodingerlikeequationr}), we can give the approximate form of the wave function at infinity:
\begin{eqnarray}
	F_{\text{L}n,\text{R}n}(t,z)
	\sim \text{e}^{\ii\omega_n( t-\frac{m_n}{\omega_n}z)}.
\end{eqnarray}
So the wave speed is $v=\frac{m_n}{\omega_n}=\frac{m_n}{\sqrt{m_n^2+a_n^2}}$, where the dispersion relation is $\omega_n^2=m_n^2+a_n^2$. It can be see that the value of $a_n$ and $m_n$ determines the velocity of the wave moving outward along the extra dimension at infinity. If $a_{n}>0$, these KK particles travel along the extra dimension at the speed slower than light speed at extra dimensional infinity.

\begin{figure}[htbp]
	\centering
	\includegraphics[width=0.45\textwidth]{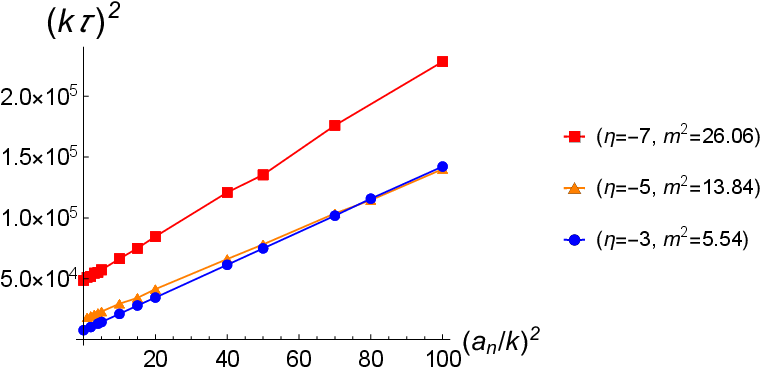}
	\vskip -4mm \caption{The relation between the lifetime of the resonance and $a_n^2$ with $q=2$.}
	\label{linearfitting}
\end{figure}

We plot the relation between the lifetime of the resonance and the parameter $a_n$ in Fig.~\ref{linearfitting}. The plot shows that the lifetime of the resonance increases with the parameter $a_n$, indicating that when the four-dimensional effective mass $m_n$ of the KK mode is a constant, the momentum of the KK mode has a positive correlation with its lifetime. Specifically, as $a_{n}$ becomes sufficiently large, the lifetime of the resonance increases approximately linearly with $a_{n}$. Thus, when the momentum on the brane or the parameter $q$ is much larger than 1 or the coupling is strong enough, the lifetime of the fermion resonance can be extremely long, possibly even lasting until the age of the universe. For example, when the five-dimensional fundamental mass $M_5$ is 10 TeV and the parameters are set to $\eta=-7, q=2,$ and $a_n=0$, the lifetime of the first resonance state is about 40 million years. This suggests that it could offer the possibility of a new class of detectable fermion KK modes with different masses.
\begin{table*}[htpb!]
	\begin{center}
		\begin{tabular}{l|cccc|cc}
			\hline
			$\,\quad\eta$       &  $\quad\text{parity}\quad$ & $m^{2}_{n}/k^2$  & $m_{n}/k$ & $P_\text{L,R}$ & $k\tau_{\text{FWHM}}$  & $k\tau_{\text{Numerical}}$
			\\ \hline
			& odd    & $\quad 2.2334\quad $ & $\quad 1.4945\quad $  & $\quad 0.9834\quad $ & $124.5$ & $859.8$
			\\
			\raisebox{2.3ex}[0pt]{$\quad-3\quad\,$} & even   & $5.5406$  & $2.3654$ & $0.5160$ & $8.01$ & $61.70$
			\\
			\hline
			& odd    & $3.2301$ & $1.7973$ & $0.9996$ & $8398.4$& $\quad 4.809\times10^{6}\quad$
			\\
			$\quad-5\quad\,$ & even   & $9.0202$ & $3.0033$ & $0.9944$ &  $202.3$ & $4864$
			\\
			& odd    & $13.8446$ & $3.7236$ & $0.7899$ & $19.07$& $118.4$
			\\
			\hline
			& odd    & $4.0289$ & $2.0072$ & $0.9999$ & $6.677\times10^6$ & $1.997\times 10^{12}$
			\\
			& even   & $11.6926$ & $3.4195$ & $0.9998$ & $1.005\times10^4$ & $7.668\times10^6 $
			\\
			$\quad-7\quad\,$ & odd    & $19.0650$ & $4.3664$ & $0.9979$ & $464.5$ & $1.758\times 10^{4}$
			\\
			& even   & $26.0595$ & $5.1059$ & $0.9296$ & $40.35$ & $353.8$
			\\
			&odd     & $31.7044$ & $5.6307$ & $0.4553$ & $6.306$ & $78.20$
			\\
			\hline
		\end{tabular}
		\caption{The mass spectra $m^2_n$ and $m_n$, relative probability $P$, and lifetime $\tau$ of the left- and right-chiral KK fermion resonances for the Yukawa coupling. $\tau_{\text{FWHM}}$ and  $\tau_{\text{Numerical}}$ are calculated by  the full width at half maximum and evolution, respectively. The parameter is set to be $q=2$. The symbols $\text{L}$ and $\text{R}$ are short for left-chiral and right-chiral, respectively.}
		\label{TableSpectraYukawa1}
	\end{center}
\end{table*}

\begin{table*}[htbp]
	\begin{center}
		\begin{tabular}{ c | c | c | c | c | c | c | c | c | c }
			\hline
			
			$\eta$   & -3 & -3.5  &-4 & -4.5 &-5 &-5.5  &-6 &-6.5 &-7        \\ \cline{2-10}
			\hline
			$k\tau$ & $61.70$ & $107.8$ & $312.1$ & $749.7$ & $4864$ & $1.940\times10^4$  & $1.363\times10^5 $  & $7.982\times10^5 $ & $7.668\times10^6 $   \\ \cline{2-10}\cline{2-10}  \hline
		\end{tabular}\\
		\caption{The lifetime $\tau$ of the second resonance.}
		\label{TableSpectraYukawa2}
	\end{center}
\end{table*}

\section{Conclusion}~\label{conclusion}

In this work, we investigated the evolution of Dirac KK modes in the thick brane. Starting from the brane solution given in Ref.~\cite{Afonso:2006gi}, we obtained the evolution equations~(\ref{schrodingerlikeequationr}) and Schr\"odinger-like equations~(\ref{schrodingerlikeequationl}).  Based on these equations, we obtained the solution of the resonances and studied their evolution. We found that the lifetime of these Dirac resonances can be very long. The resonance is a kind of important and interesting object in the braneworld. However, the dynamical behavior of Dirac resonances in the braneworld model has not been fully investigated. Thus, we still lack a comprehensive and profound understanding of the nature of the Dirac KK mode in braneworld model.

The fermion resonances were derived with the relative probability method established in Ref.~\cite{Liu:2009ve}. We found that the number of the resonances and the relative probability of the first resonance increase with the absolute value of the coupling parameter $|\eta|$ and the parameter $q$, which can be seen from Fig.~\ref{Figprobabilityq}, Fig.~\ref{Figwavefuction}, and Tab.~\ref{TableSpectraYukawa2}.  Such behavior is similar with the previous literature~\cite{Zhang:2016ksq,Xie:2019jkq,Liu:2009ve,Almeida:2009jc}.  We also observed that larger relative probability corresponds to longer lifetime of the resonance, which is confirmed by the evolution of the KK mode presented in Figs.~\ref{Figenenrgyevolution} and ~\ref{Fixzevolution}. Additionally, we investigated the evolution of the nonresonance. We found that its amplitude and energy decay promptly at the beginning. However, the later damping occurs in a similar manner as the resonances. Our results suggest that the nonresonance evolves into a superposition of the resonances, as demonstrated in Fig.~\ref{Fourierdecomposition}. Furthermore, we obtained that the lifetime of a resonance almost linearly increases with the value of the three-dimensional momentum $a_n$, as shown in Fig.~\ref{linearfitting}. If the parameter $a_n$, $q$, or $|\eta|$ is large enough, the resonance may have a very long lifetime on the brane. It could offer the possibility of a new class of detectable fermions with different masses. At the same time, the measurement of the fermion mass can also limit the parameters in the model.

Next, it would be worthwhile to investigate whether the similar technique could be applied to the gravitational perturbation in extra dimensional theory, potentially identifying it as dark matter candidate. The effective potential in such theories typically has multiple potential barriers, which could allow for the study of the gravitational echo phenomenon in extra dimensions, analogous to that in black hole theory.

\section*{Acknowledgments}
We are thankful to J.~Chen and J.-J.~Wan for useful discussions. This work was supported by the National Key Research and Development Program of China (Grant No. 2020YFC2201503), the National Natural Science Foundation of China (Grants No.~12105126, No.~11875151, and No.~12247101), the 111 Project under (Grant No. B20063), the Fundamental Research Funds for the Central Universities (Grant No. lzujbky2021-pd08), the China Postdoctoral Science Foundation (Grant No. 2021M701531), and ``Lanzhou City's scientific research funding subsidy to Lanzhou University".

\appendix
\section{Convergence test}

In the appendix, we will briefly discuss the convergence of our simulations, for which we compare the energy by using three space steps $h=1/32, 1/40, 1/50$. This energy belongs to the first resonance with $\eta=-3$.

\begin{figure}[htbp]
	\centering
	\includegraphics[width=0.33\textwidth]{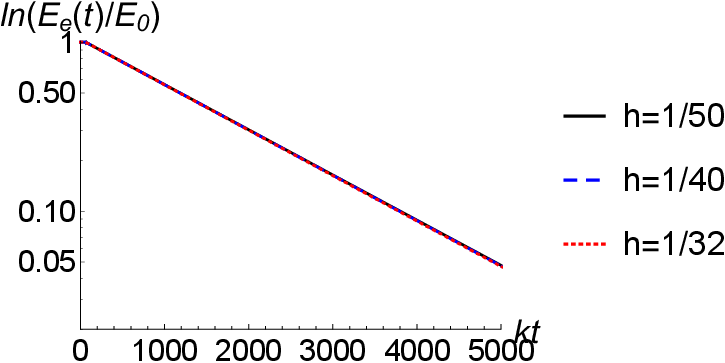}		
	\vskip -4mm \caption{The variation of the first resonance energy decays with with the space step $h$ for the case $\eta=-3$.}
	\label{stepchange}
\end{figure}

In general, the numerical solutions of a system converge according to the following rule
\begin{equation}
	\phi_h-\phi\propto h^n,
\end{equation}
\begin{figure}[htbp]
	\centering
	\includegraphics[width=0.29\textwidth]{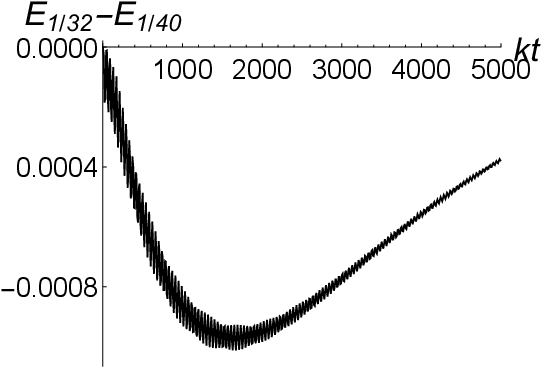}
	\includegraphics[width=0.29\textwidth]{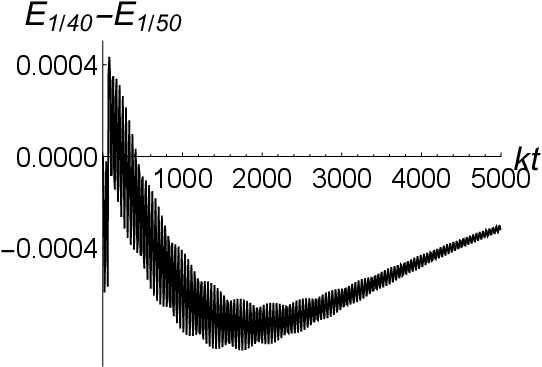}
	\vskip -4mm \caption{Plot of the energies error with different spatial step $h$.}
	\label{energyerror}
\end{figure}
where $h$ is the space step and $n$ is the convergence order. We use the convergence factor $Q$ to study its convergence behavior. Its form is
\begin{equation}
	Q=\frac{\phi_1-\phi_2}{\phi_2-\phi_3}=\frac{h_1^n-h_2^n}{h_2^n-h_3^n}.
	\label{convergencefactor}
\end{equation}

\begin{figure}[htbp]
	\centering
	\includegraphics[width=0.29\textwidth]{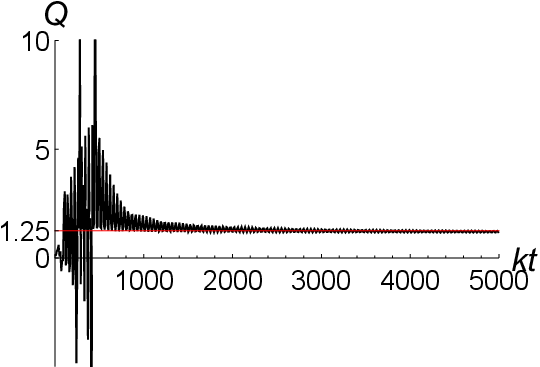}
	\vskip -4mm \caption{Plot of the evolution of the convergence factor $Q$.}
	\label{convergencefactorfig}
\end{figure}

We give the corresponding results in Figs.~\ref{energyerror} and \ref{convergencefactorfig}. Using these results and the relation (\ref{convergencefactor}), we plot the convergence factor $Q$ as a function of time in Fig.~\ref{convergencefactorfig}. We can see that $Q \simeq 1.25$, such value of $Q$ means that the convergence order is of about 1. It can also be seen from Fig.~\ref{energyerror} that as the space step size decreases, the energy error of the asynchronous step also decreases.

\bibliography{ref}

\end{document}